# A Soft Computing technique in rainfall forecasting

Surajit Chattopadhyay[1] and Manojit Chattopadhyay[2]
Pailan College of Management and Technology
Kolkata –700 104
email:[1] surajit_2008@yahoo.co.in  [2] chattomanojit@mail.yahoo.com

*Abstract: Soft Computing techniques have opened up new avenues to the forecasters of complex systems. Atmosphere is a complex system and all the atmospheric parameters carry different degrees of complexity within themselves. Endeavor of the present research is to adopt Artificial Neural Network as a Soft Computing technique to anticipate the average monsoon rainfall over India. After a rigorous study, the said Neural Net technique has been found adroit and results have been compared with those obtained through conventional techniques.*

## 1. INTRODUCTION

The Soft Computing techniques are based on the information processing in biological systems. The complex biological information processing system enables the human beings to survive with accomplishing tasks like recognition of surrounding, making prediction, planning, and acting accordingly.

For a computing system to have human like information processing facility, it should be flexible enough to support three features: openness, robustness, and real time processing. Openness of a system is its ability to adapt or extend on its own to cope with changes encountered in the real world. Robustness of a system means its stability and tolerability when confronted with distorted, incomplete, or imprecise information. The real time characteristic implies the ability of the system to react within a reasonable time in response to an event. Information processing systems with all these three characteristics are known as real world computing (RWC) systems [1]. A RWC system should, therefore, be capable of distributed representation of information, massively parallel processing, learning, and self-organization in order to achieve flexibility in information processing. Thus, Soft Computing can be viewed as the key ingredient of RWC systems. Several authors (e.g. [2]; [3]; [4]) have discussed the potentials of the soft computing techniques in solving real world problems. Soft Computing has three basic components, namely, Artificial Neural Network, Fuzzy Logic, and Genetic Algorithm.

Present paper adopted Artificial Neural Network as predictive algorithm for monsoon rainfall over India.

## Artificial Neural Network in forecasting atmospheric events- an overview

ANNs have recently become important alternative tool to conventional methods in modelling complex non-linear relationships. In the recent past, the ANN has been applied to model large data with large dimensionality.

Hu (1964) [5] initiated the implementation of Artificial Neural Network, an important Soft Computing methodology in weather forecasting. McCann (1992) [6] developed Artificial Neural Network models to give 3-7 hr forecast of significant thunderstorms on the basis of surface based lifted index and surface moisture convergence. The two neural networks produced by them were combined operationally at National Severe Storms Forecast Center, Kansas City, Missouri to produce a single hourly product and was found to enhance the pattern recognition skill. Cook and Wolfe (1991) [7] developed a neural network to predict average air temperatures. Zhang and Scofield (1994) [3] applied Artificial Neural Network (ANN) in prediction of rainfall through satellite data.

## Data used in the present paper

In India, the months June, July, and August are identified as the summer-monsoon months. Thus, the present study explores the data of these three months corresponding to the years 1871-1999. This paper develops ANN model step-by-step to predict the average rainfall over India during summer- monsoon by exploring the data available at the website http://www.tropmet.res.in published by Indian Institute of Tropical Meteorology.

## 2. IMPLEMENTATION OF MULTILAYER PERCEPTRON

Feed forward neural network, also known as Multilayer Perceptron (MLP), has become popular tool for solving complex prediction as well as classification problems.  In general, a Multi Layer Perceptron consists of a lowermost input layer, any number of hidden layers, and an output layer at the top. In a network, the total input received by neuron 'j' in $(h+1)$ th layer is defined as;

$$X_j^{h+1} = \sum_i Y_i^h W_{ji}^h - \theta_j^{h+1} \qquad (1)$$

Where, $Y_i^h \rightarrow$ State of the $i^{th}$ neuron in the preceding $h^{th}$ layer,

$W_{ji}^h \rightarrow$ Weight of the connection between $i^{th}$ neuron of $h^{th}$ layer and $j^{th}$ neuron of $(h+1)^{th}$ layer,

$\theta_j^{h+1} \rightarrow$ Threshold of $j^{th}$ neuron in $(h+1)$ layer.

For convenience, the threshold '$\theta$' is taken to be zero.

Equation (1), thus, modifies to;

$$X_j^{h+1} = \sum_i Y_i^h W_{ji}^h \qquad (2)$$

Where

$$Y_i^h = [1/\{1 + \exp(-X_j^h)\}]$$

For $h > 0$

$\qquad = X_j^0$

For h=0

The values $W_{ji}^h$ are random values within [–0.5, 0.5]. Perceptron of zero thresholds is more convenient to deal with. This corresponds to linear separations that are forced to pass through the origin of the input space.

To develop the predictive model, the monsoon months' (June-August) data of year $y$ have been used to predict the average monsoon rainfall of year ($y+1$). First 75% of the available data are used as training set and the remaining 25% are used as the test set.

The multilayer perceptron model is framed with the data set divided into test and training cases. The test cases are arbitrarily chosen for the available dataset so as to maintain the generality. The model is trained up to 50 epochs. The learning rate parameter is fixed at 0.4 and the momentum rate is chosen 0.9. The weight matrix is framed with the values between –0.5 and 0.5. Least mean squared error is chosen as the stopping criteria for the learning procedure. In the present Multilayer Perceptron, 2 hidden layers are chosen and each hidden layer contains 2 nodes. A sequential learning procedure is adopted to train the perceptron. The final weight matrix is found to be

| | | | | |
|---|---|---|---|---|
| **Hdn1_bias** | 0.0000 | 0.0000 | 0.0000 | 0.0000 |
| **Hdn1_Nrn1** | -0.5563 | -0.5496 | -0.9791 | -0.5501 |
| **Hdn1_Nrn2** | -0.5081 | -0.5560 | -0.9159 | -0.5709 |
| | 1.0000 | 0.5081 | 0.5171 | |
| **Op_bias** | 0.0000 | 0.0000 | 0.0000 | 0.0000 |
| **Op_Nrn1** | 0.2175 | 0.9678 | 0.9046 | 1.1770 |
| | 1.0000 | 0.7644 | | |

After training the perceptron, the model is tested with the chosen validation or test set. The predicted values and the associated predictor values are displayed in table1. To make the model output more understandable, the model outputs and the target outputs for the test cases are displayed in the figure 1.

## 3. PERFORMANCE EVALUATION

Performance of the model is evaluated through computation of overall prediction error (PE) defined as

$$PE = \frac{\langle |y_{predicted} - y_{actual}| \rangle}{\langle y_{actual} \rangle} \times 100 \qquad (3)$$

Where, $\langle \; \rangle$ implies average over the test cases.

In the present case, PE is 10.2%, that is, a small overall prediction error is being yielded by the perceptron model. Percentage errors of prediction in the individual test cases are displayed in figure 2. It is found that in 87.5% test cases, the prediction error is less than 20%. Thus, there is a high prediction yield.

Now, the performance of the neural net model is compared with conventional persistence forecast. It is found through equation (3), that PE in the case of persistence forecast is 18.3%. Therefore, Neural Net, in the form of Multilayer Perceptron is found to be adroit in the prediction of monsoon rainfall over India.

## 4. CONCLUSION

Present study concludes that Soft Computing as Artificial Neural Network can be of great use in prediction monsoon rainfall over India. If more input parameters are available, then a prediction of higher accuracy would be possible.

## 5. ACKNOWLEDGE

The authors wish to express sincere thanks to Professor Souvik Roy, Head of the Department of Information Technology of Pailan College of Management and Technology for his continuous inspiration towards this work.

| Test cases | June (y) | July (y) | August (y) | Actual average rainfall (y+1) | Prediction through ANN (y+1) |
|---|---|---|---|---|---|
| 12 | 208 | 365 | 128.1 | 190.5666667 | 204.4048773 |
| 14 | 136 | 350.1 | 225.8 | 215.0666667 | 204.6750073 |
| 17 | 148.3 | 341.7 | 249.3 | 198.6 | 204.2476912 |
| 27 | 86 | 246.6 | 276 | 207.3333333 | 207.736091 |
| 29 | 150 | 94.5 | 88.8 | 202.8 | 215.8517001 |
| 30 | 70.5 | 232.3 | 305.6 | 180.8666667 | 207.9729845 |
| 31 | 78.5 | 215.5 | 248.6 | 153.6 | 209.5307804 |
| 35 | 49.5 | 225.3 | 123.3 | 208.3666667 | 213.1694524 |
| 36 | 162.1 | 271 | 192 | 210.9666667 | 206.7439003 |
| 41 | 139.1 | 120.4 | 182.3 | 207.8333333 | 212.6632418 |
| 45 | 127.5 | 185.8 | 163 | 226.1666667 | 211.0957005 |
| 46 | 158.6 | 226.6 | 293.3 | 234.1333333 | 206.1558816 |
| 55 | 167.3 | 253.8 | 169.6 | 211.3666667 | 207.5558978 |
| 62 | 75.5 | 350.2 | 168.6 | 236.7 | 207.093281 |
| 65 | 105.8 | 327.3 | 158.3 | 194.8666667 | 207.1941048 |
| 68 | 226.8 | 256 | 192.8 | 182 | 205.6046159 |
| 72 | 137 | 350.3 | 275.7 | 187 | 203.8736594 |
| 79 | 95.3 | 279.7 | 184.8 | 194.4666667 | 208.3724254 |
| 85 | 156 | 160.8 | 333.8 | 254.6333333 | 207.306592 |
| 89 | 123.8 | 357.7 | 255.3 | 199.5666667 | 204.2774935 |
| 94 | 125.5 | 267.8 | 275.7 | 162.3 | 206.1444557 |
| 95 | 67.8 | 270.3 | 148.8 | 164.1666667 | 210.2871189 |
| 100 | 196.3 | 192.6 | 330.1 | 210.7333333 | 205.5260319 |
| 104 | 72.3 | 205.3 | 203.6 | 227.5333333 | 211.13565 |
| 105 | 149.5 | 260 | 273.1 | 228.5666667 | 205.8399932 |
| 107 | 178.6 | 291.7 | 225.8 | 242.8333333 | 205.1952834 |
| 112 | 79.5 | 219.1 | 249.1 | 229.1 | 209.3700871 |
| 113 | 118.4 | 266.7 | 302.2 | 192.8333333 | 205.8583788 |
| 114 | 95.4 | 217.5 | 265.6 | 176.6 | 208.6060492 |
| 115 | 97.9 | 230.8 | 201.1 | 193.8333333 | 209.5297575 |
| 123 | 135.1 | 287 | 190.1 | 261.7666667 | 206.9964667 |
| 124 | 206.6 | 322.1 | 256.6 | 175.9333333 | 203.4527641 |

Table 1- The predictors (2<sup>nd</sup>, 3<sup>rd</sup>, 4<sup>th</sup> columns), the predictand (5<sup>th</sup> column), and the prediction (6<sup>th</sup> column).

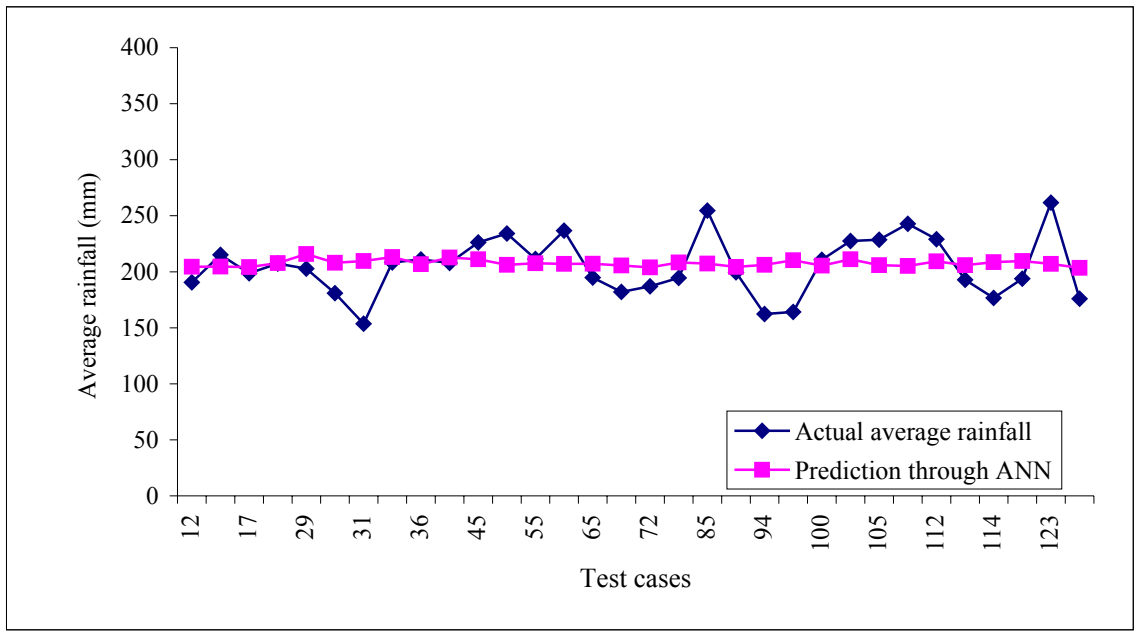

Figure 1- Schematic showing the actual average monsoon rainfall and those predicted through ANN in the test cases.

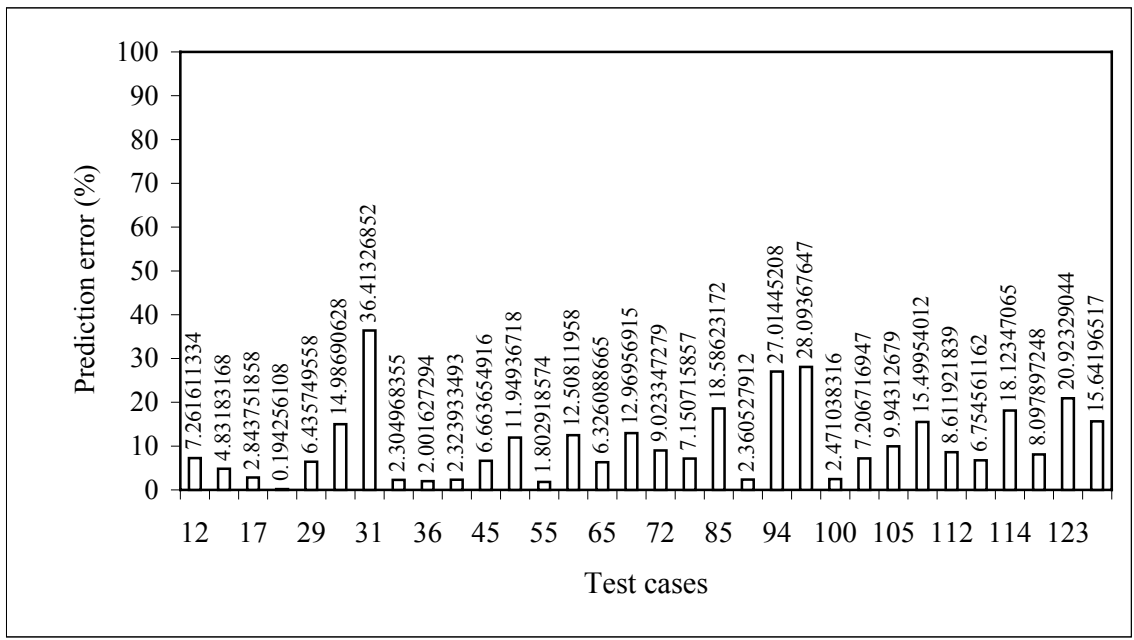

. Figure 2- Schematic showing percentage errors of prediction in the test cases.